\documentclass[aps,prl,reprint,groupedaddress,bibnotes,superscriptaddress,
preprintnumbers,amsmath,amssymb,lengthcheck]{revtex4-1}
\usepackage{amsmath,amssymb,graphicx}
\usepackage{color}

\begin{document}

\title{Real-time observation of dissipative optical soliton molecular motions}
\author{Katarzyna Krupa}\email[]{Corresponding author: katarzyna.krupa@u-bourgogne.fr}
\affiliation{Universit\'e de Bourgogne Franche-Comt\'e, ICB, UMR CNRS 6303, 9 Av. A. Savary, 21078 Dijon, France}
\author{K. Nithyanandan}
\affiliation{Universit\'e de Bourgogne Franche-Comt\'e, ICB, UMR CNRS 6303, 9 Av. A. Savary, 21078 Dijon, France}
\author{Ugo Andral}
\affiliation{Universit\'e de Bourgogne Franche-Comt\'e, ICB, UMR CNRS 6303, 9 Av. A. Savary, 21078 Dijon, France}
\author{Patrice Tchofo-Dinda}
\affiliation{Universit\'e de Bourgogne Franche-Comt\'e, ICB, UMR CNRS 6303, 9 Av. A. Savary, 21078 Dijon, France}
\author{Philippe Grelu}\email[]{philippe.grelu@u-bourgogne.fr}
\affiliation{Universit\'e de Bourgogne Franche-Comt\'e, ICB, UMR CNRS 6303, 9 Av. A. Savary, 21078 Dijon, France}

\date{\today}

\begin{abstract}
Real-time access to the internal ultrafast dynamics of complex dissipative optical systems opens new explorations of pulse-pulse interactions and dynamic patterns. We present the first direct experimental evidence of the internal motion of a dissipative optical soliton molecule generated in a passively mode-locked erbium-doped fiber laser. 
We map the internal motions of a soliton pair molecule by using a dispersive Fourier-transform imaging technique, revealing different categories of internal pulsations, including vibration-like and phase drifting dynamics. Our experiments agree well with numerical predictions and bring insights to the analogy between self-organized states of lights and states of the matter. 
\end{abstract}


\maketitle


The soliton is a universal concept applicable to a large class of solitary wave propagation effects that can be observed in most branches of nonlinear science, from fluid dynamics and biology to plasma physics and photonics \cite{Dau15}. Due to the progress in optical materials and laser technologies, as well as the relative simplicity of optical schemes, solitons in nonlinear optics have been the subject of constant research, and constitute a fertile land for promoting analogies into other scientific areas \cite{Kiv03}. 
From the 1990s, the concept of optical solitary waves was progressively extended to propagation media including gain and loss, namely dissipative systems \cite{Ros02,Akh05,Gre16}. Dissipative solitons in photonics can be defined as localized formations of an electromagnetic field that are balanced through the energy exchange with the environment in presence of nonlinearity, dispersion and/or diffraction. In addition to answering a span of real world applications, echoing the deployment of fiber amplifiers and the acceleration of fiber laser research, dissipative solitons emerged as a new soliton concept associated with distinct attributes \cite{Akh05,Gre12}. Once given the parameters of the propagation medium, dissipative solitons will generally be endowed with a fixed profile. This feature is of great practical interest for the design of stable pulse sources such as mode locked lasers. Reciprocally, mode locked lasers constitute a powerful test-bed for the investigation of dissipative soliton dynamics \cite{Gre12}. 

In the frame of nonlinear dynamics, the fixed profile of a dissipative soliton results from the existence of a fixed-point attractor. Besides single solitons, multisoliton patterns \cite{Mal91,Akh97} can also be formed, when the rate of energy supply into the system is increased. As a matter of fact, multiple pulsing is a common behavior among passively mode-locked fiber lasers \cite{Nak91}. When several dissipative solitons coexist in a laser cavity, they will always interact, considering the variety of interactions schemes and the virtually infinite time to manifest them. For instance, incoherent interactions based on gain depletion and recovery are weakly repulsive, and promote a regular distribution of the soliton pulses along the cavity that multiplies the laser repetition rate, namely harmonic mode locking \cite{Kut98, Lec13}. Coherent interactions are based on pulse overlap, and can bring a strong attractive contribution. In specific parameter ranges, these interactions lead to the self-assembly of stable multisoliton bound states \cite{Mal91, Akh97}, also termed soliton molecules \cite{Str05,Gre12}. Indeed, the analogy with matter molecules is, by certain aspects, particularly striking. Soliton molecules can form or dissociate \cite{Roy05}. They can exist in various isomers, considering the relative temporal separation and phase difference between adjacent pulses as internal dynamical degrees of freedom \cite{Leb06}. While soliton pairs constitute the central soliton molecule case \cite{Tan01,Gre02}, large solitons macromolecules and crystals comprising $\sim10^{3}$ dissipative solitons can also self-assemble \cite{Hab08}. 

From the applied perspective, soliton molecules are being considered as a way to provide robust upper bits for multilevel encoding in optical communications \cite{Str05}. From the point of view of fundamental physics, it is fascinating to explore how far the analogy with matter molecules could be developed. In this respect, vibrations and pulsations constitute vivid possible expressions of the dynamics of dissipative soliton molecules. Vibrating soliton pairs have been theoretically investigated: in principle, they result from a Hopf-type bifurcation that triggers an oscillation of the internal degrees of freedom of the soliton pair molecule \cite{Gra06,Sot07,Zav09}. The period of oscillation, albeit significantly larger than the cavity round trip time, remains small compared to the  acquisition time of the spectral and autocorrelation measurements that have been used so far to characterize the ultrafast  dynamics of soliton molecules. In previous ultrafast fiber laser experiments, indirect signatures based on time-averaged experimental measurements have been attributed to vibrating soliton pairs \cite{Gra06,Ort10,Li14,Wan16}. However, the proof of existence of soliton molecular motions is still lacking to date, and calls for advanced ultrafast characterization approaches.

The development of advanced methods to visualize, in real time, ultrafast processes occurring in chemical or biological structures is attracting great attention \cite{Chem16}. In photonics, the dispersive Fourier-transform (DFT) measurement technique has recently permitted to experimentally resolve complex ultrafast nonlinear phenomena, from stochastic fluctuations in modulation instability to optical rogue waves and soliton explosions in ultrafast fiber lasers \cite{Sol12, Lec14, Run15}. The principle of the DFT technique consists of stretching a train of optical pulses in a dispersive medium that cumulates a group-velocity dispersion large enough to map the spectrum of each optical pulse into a temporal waveform, which is a conceptual analogue to the far-field limit in paraxial diffraction \cite{God13}. 

In this article, we provide the first unambiguous experimental measurements that prove the existence of internal motions within soliton pair molecules. Based on a relatively simple yet powerful DFT technique, our measurements reveal the evolution of the relative phase and/or temporal separation between the two bound solitons, along each cavity round trip, allowing to visualise these internal soliton molecular motions in real time. We highlight two markedly different categories of soliton molecular dynamics. The first one is a combined oscillation of both relative soliton phase and temporal separation, which makes a vivid analogy with the vibrations of a matter diatomic molecule. The second one mainly consists of a sliding relative phase, and is therefore a distinct feature of optical soliton molecules. Both experimental cases are supported by detailed numerical simulations of the fiber laser ultrafast dynamics. 

\begin{figure}
\includegraphics[width=\linewidth]{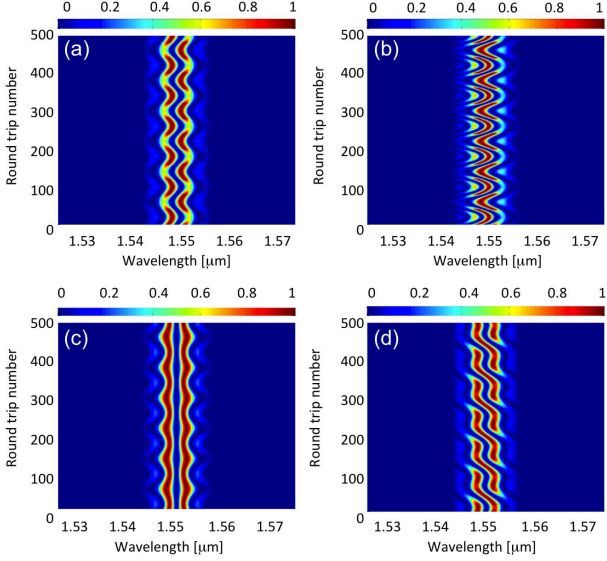}
\caption{Analytical evolution of spectral intensity profile versus round trips for two-pulses co-propagated with: 
(a) oscillating phase ($\phi_{0}= \pi$, $A_{\phi}=0.15\pi$), 
(b) oscillating phase and separation ($\phi_{0}= \pi$, $A_{\phi}=0.5\pi$, $A_{\tau}=\tau_{0}/4$),
(c) oscillating temporal separation ($A_{\tau}=\tau_{0}/4$), 
(d) diverging sliding phase ($\phi_{0}= \pi$, $A_{\phi}=0.15\pi$). Pulse duration: 0.6 ps, average temporal separation $\tau_{0}$=1 ps.} \label{fig:1}
\end{figure}

\begin{figure}
\includegraphics[width=\linewidth]{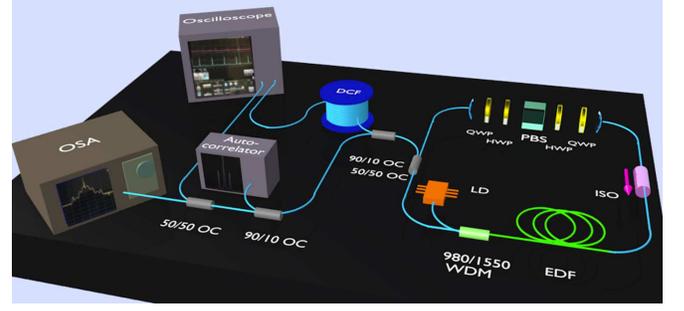}
\caption{Experimental setup of NPE-based mode-locked fiber laser. OC - optical coupler, WDM - wavelength division multiplexer, PBS - polarizing beam splitter, ISO - polarization-independent isolator, LD - 980-nm laser diode, 
HWP and QWP - half-waveplate and quater-waveplate, EDF - erbium-doped fiber, SMF - standard single mode fiber (blue lines).} \label{fig:2}
\end{figure}

Since our characterization of the soliton molecular motions is based on the analysis of successive, shot-to-shot, optical spectra, it is instructive to present the correspondence between the internal degrees of freedom of the soliton molecule and the distinct spectral features. 
The soliton pair molecule consists of two bound solitons, whose temporal separation $\tau$ and relative phase $\phi$ may oscillate with the propagation distance $z$ -- or equivalently, the number of cavity round trips. Let us consider a simple example of the sum of two identical Gaussian pulses, $s(t)=\psi\left(t+\frac{\tau}{2}\right)+\psi\left(t-\frac{\tau}{2}\right)e^{-i\phi}$, with  $\psi$ being a Gaussian envelope and $t$ the time in the co-moving reference frame. 
The spectral intensity $I(\omega)$ to be recorded will be:
\begin{eqnarray}
I(\omega)=|S(\omega)|^2=4|\Psi(\omega)|^2\cos^2\left(\omega \frac{\tau}{2}+\frac{\phi}{2}\right),
\label{gnlse}
\end{eqnarray}
where $\Psi(\omega)$ and $S(\omega)$ are the Fourier transforms of $\psi(t)$ and $s(t)$, respectively.
The stationary soliton pair molecule is obtained when $\tau$ and $\phi$ are constant. This leads to fixed spectral fringes of high contrast that can be recorded by a conventional optical spectrum analyzer (OSA), through averaged measurements, the interfringe being inversely proportional to the temporal separation \cite{Tan01,Gre02}. 

When considering $\tau(z)=\tau_{0}+A_{\tau}sin(z)$ and/or $\phi(z)=\phi_{0}+A_{\phi}sin(z)$, we model a harmonically vibrating soliton molecule instead. The oscillation in $\phi�$ produces an overall oscillatory sliding of the spectral fringes (see Fig.\ref{fig:1}(a)), whereas that of $\tau$ entails a breathing of the spectral interfringe, as show in Fig.\ref{fig:1}(c). The result of combined oscillations ($\tau$ and $\phi$) is displayed on Fig.\ref{fig:1}(b). We also illustrate a case when the relative phase is infinitely drifting away, with $\phi(z)=-z+(\phi_{0}+A_{\phi}sin(z)$), see Fig.\ref{fig:1}(d). 
Observing these effects requires an advanced real-time spectral characterization otherwise, only faint indirect information such as the blurring of spectral fringes, will hint to a non-stationary dynamics \cite{Gra06, Ort10, Wan16}. 

To experimentally investigate soliton molecule dynamics, we built a mode-locked Erbium-doped fiber ring laser emitting at $\lambda\simeq$ 1.55$\mu$m, sketched on Fig.\ref{fig:2}. The laser cavity incorporates a 0.55-m long Erbium-doped silica fiber (EDF, 110 dB/m absorption at 1530 nm, 4 $\mu$m core diameter), which is backward-pumped by a 980 nm laser diode. The other fibers are standard single-mode fibers (SMF-28), with a total length of 3.45 m. The group velocity dispersion at 1.55 $\mu$m is +13.5 ps$^{2}$.km$^{-1}$ for the EDF and  -22.9 ps$^{2}$.km$^{-1}$ for the SMF, yielding an anomalous path-averaged cavity dispersion
$\beta_{2}$=-17.8 ps$^{2}$.km$^{-1}$. 
To mode-lock the fiber laser, we use nonlinear polarization evolution (NPE), which takes place in the fibers and is discriminated by the intracavity polarizer (PBS) \cite{Gre02}. NPE provides a virtual, quasi-instantaneous, saturable absorber effect whose transfer function is readily tuned by adjusting the orientations of the waveplates. As a consequence, a wide range of ultrafast dynamics can be accessed in a reproducible way. 
The total cavity length, 4.43 m entails a fundamental repetition rate of 47.95 MHz. 
The experimental characterization, performed through the 50/50 fiber output coupler, is detailed hereafter. 
By appropriate adjustments of the waveplates, we achieved stable mode-locking for a pump power $P_{P}$ in the range 90-380 mW, and could select either single or multiple soliton operation. 
\begin{figure}
\includegraphics[width=\linewidth]{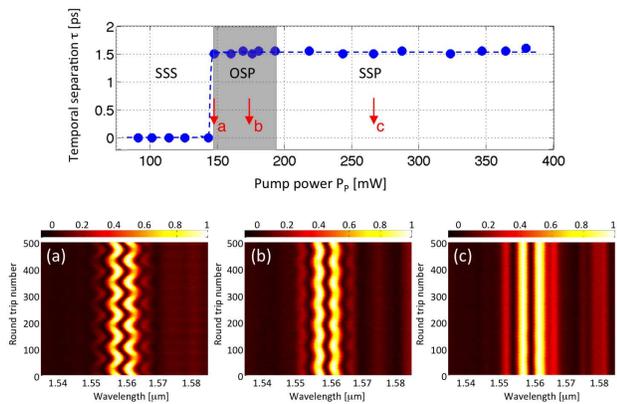}
\caption{Top panel: Experimental examples of two transitions from single soliton (SS) to soliton pair (SP), and from stationary soliton pair (SSP) to oscillating soliton pair (OSP), which occur with a change of pump power ($P_{P}$). SSS: stationary single soliton. Bottom panels: Corresponding 2D contour plots of the sequence of 500 consecutive single-shot laser spectra for (a) $P_{P}$=147 mW, (b) $P_{P}$=175 mW, and (c) $P_{P}$=265 mW.} \label{fig:3}
\end{figure}
\begin{figure}
\includegraphics[width=\linewidth]{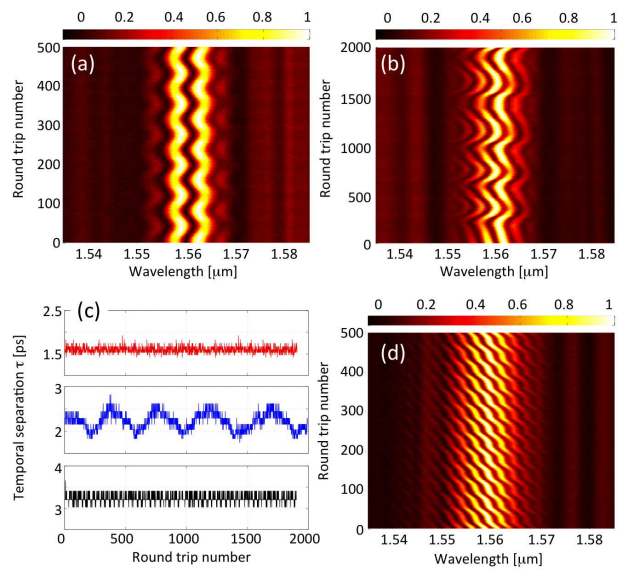}
\caption{Experimental results of a soliton pair molecule with (a and red curve) oscillating phase; (b and blue curve) vibrating molecule ; (c and black curve) diverging sliding phase. Panels a, b and d: 2D contour plot of consecutive single-shot laser spectra. Panel c: Temporal separation between two pulses $\tau$ as a function of round trips.} \label{fig:4}
\end{figure}

To perform DFT real-time measurements of the shot-to-shot spectral evolution, the laser output was 
propagated through a 1345-m long normally-dispersive fiber link (DCF), 
providing a total accumulated dispersion of 163~ps$^{2}$. The temporal optical intensity was detected by a high-speed 45-GHz photodiode and recorded by a 6-GHz, 40-GSa/s real-time oscilloscope. 
This led to an electronic-based spectral resolution of 0.16 THz (1.3 nm), which was suitable to analyze ultrafast dynamics having a spectral span exceeding 10 nm. 
In the following, we focus on oscillating soliton pair molecules generated with a FWHM duration of 0.66 ps and a temporal separation in the range of 1.5-3.3 ps. The two-pulse ultrafast regime was confirmed by multishot second-order optical autocorrelation. 

As discussed in the introduction, the onset of soliton molecule vibrations results from a Hopf-type bifurcation. 
As shown in the top panel of Fig.\ref{fig:3}, the increase of the pump power above 144 mW produced an abrupt transition from a single soliton (SS) to a soliton pair (SP).  We further analyzed the transition between a stationary (phase-locked) soliton pair (SSP) and an oscillating soliton pair (OSP), which manifested as a supercritical bifurcation when the pump power was decreased. Below $P_{P}$=193 mW, the bifurcation occurs and the phase difference between the two solitons starts to periodically oscillate with a continuously growing amplitude $A_{\phi}$ as $P_{P}$ decreases, until the transition back to stationary single pulse (SSS) dynamics takes place. Bottom panels of Fig.\ref{fig:3} illustrate the spectral evolution as a function of cavity round trips, where the first two examples are placed in the OSP regime, and the last one in SSP regime. 

%
The large diversity of bifurcation scenarii and modes of oscillating soliton molecules strongly depends on the polarization controller settings. We illustrate the later by three different recordings shown in Fig.\ref{fig:4}, where the panels (a), (b) and (d) display 2D contour plots of hundreds of consecutive single-shot experimental spectra. 
%
Sharing similar features with Fig.\ref{fig:1}(b), Fig.\ref{fig:4}(b) presents the dynamics of a vibrating soliton molecule that combines the oscillations of both $\phi$ and $\tau$, here with the same period of 430 round trips. The retrieved intra-molecular temporal separation is displayed as the oscillating blue curve in Fig.\ref{fig:4}(c). 
Figure \ref{fig:4}(a) reveals the existence of a soliton molecular motion dominated by the oscillation of the relative phase, with a period of 80 cavity round trips -- no oscillation of the intra-molecular separation could be evidenced in this case, also similar to the OSP dynamics displayed in Fig.\ref{fig:3}. These classes of soliton molecular dynamics were numerically predicted  in Refs.\cite{Gra06, Sot07,Zav09}. 
As a contrasting case, we also prove the existence of a dynamics characterized by the infinite divergence of the relative phase, similar to that illustrated in Fig.\ref{fig:1}(d). The spectral measurements are presented in Fig.\ref{fig:4}(d), and the analysis reveals a diverging relative phase without any significant modification of the temporal separation (see black curve in Fig.\ref{fig:4}(c)). Such diverging phase dynamics was predicted in Ref.\cite{Zav09} and inferred from Ytterbium fiber laser experiments \cite{Ort10}, albeit without the confirmation from real-time measurements. 

%
\begin{figure}
\includegraphics[width=\linewidth]{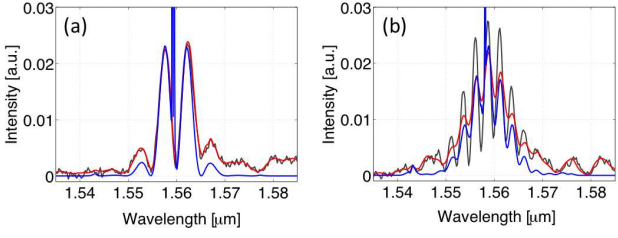}
\caption{Comparison between the average spectrum recorded by the OSA (blue curves) and the average spectrum calculated from 1000 single-shot spectra (red curves), in the case of (a) a stationary soliton pair and (b) a soliton pair molecule with diverging spliding phase. Example of a single-shot spectrum is provided grey curves.} \label{fig:5}
\end{figure}
\begin{figure}
\includegraphics[width=\linewidth]{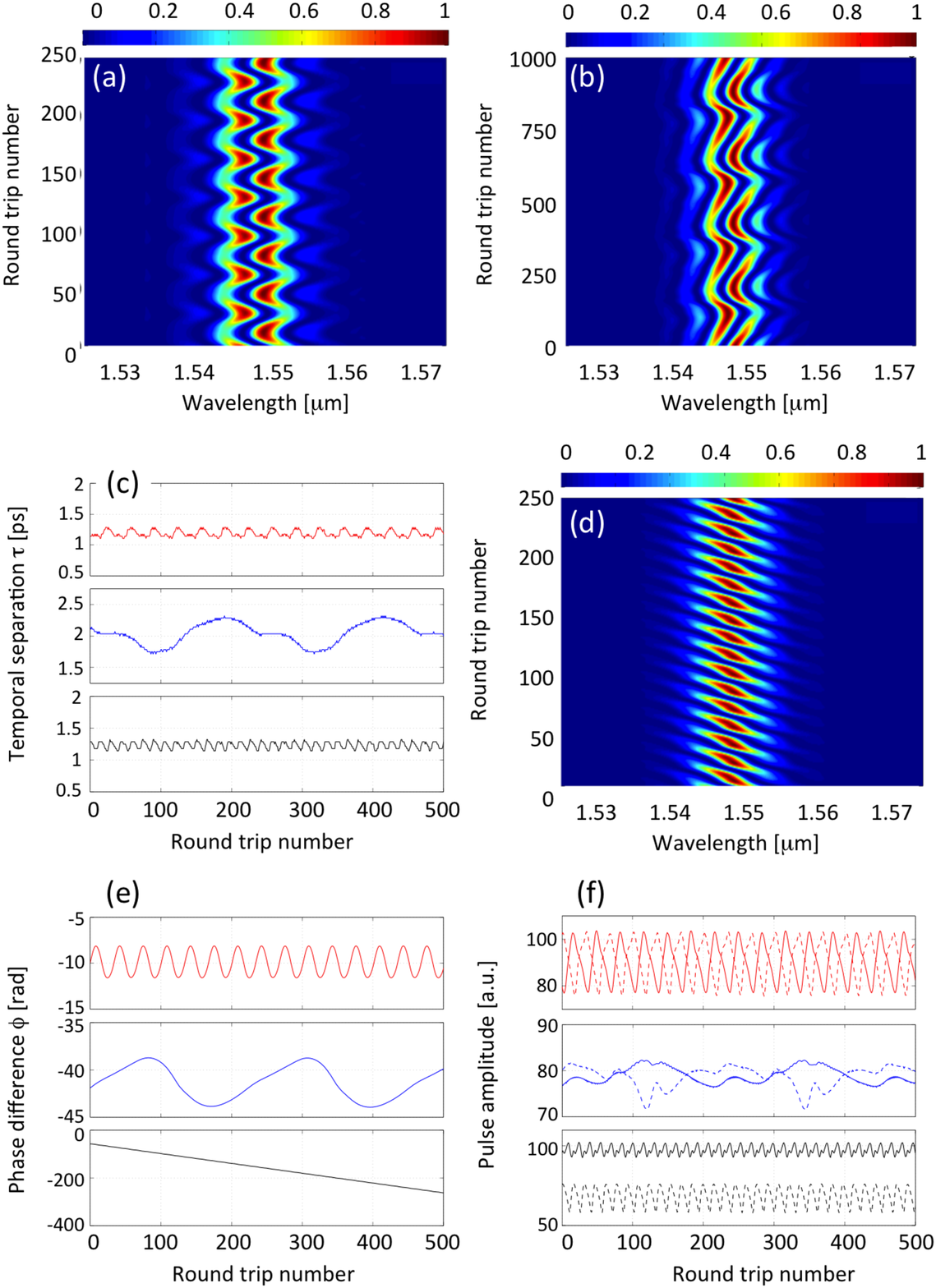}
\caption{Numerical simulations of a soliton pair molecule with (a, and red curves) oscillating phase ; (b and blue curves)  vibration dynamics ; (d and black curves) diverging sliding phase ; (c) temporal separation $\tau$ ; (e) phase difference $\phi$ and (f) amplitude of leading and trailing pulses, as a function of cavity round trips.} \label{fig:6}
\end{figure}
%

We cross-checked the calibration of the DFT real-time spectral measurements by comparing the average of 1000 consecutive spectra, typically, with the averaged spectrum directly obtained by the OSA, see Fig.\ref{fig:5}. The agreement is satisfactory, considering the limited dynamical range resulting from the digitization of the fast oscilloscope in DFT measurements. 

In order to confirm and better interpret the complex internal dynamics of soliton molecules induced in the passively mode-locked fiber laser, we developed the following numerical model. It consists of a piece-wise propagation model, which takes into account the different fiber segments that constitute the laser cavity, while adopting a scalar effective approach to the saturable absorber effect. In the fibers, propagation follows the generalized nonlinear Schr{\"o}dinger equation (GNLSE):
\begin{eqnarray}
 \frac{\partial \psi}{\partial z} + \frac{i}{2}\beta_{2}\frac{\partial^{2} \psi}{\partial t^{2}}
 -\frac{g(z)}{2} \psi = i \gamma \left| \psi \right|^{2} \psi ,
\label{gnlse}
\end{eqnarray}
where $\gamma$ is the Kerr non-linearity coefficient, and $g$ is the gain coefficient. The latter is null for the SMF, considered lossless. In the EDF, the gain $g$ includes saturation and spectral response, as in Ref.\cite{Run14}. It is therefore z-dependent on the average signal and pumping powers, the average power dynamics being described by standard rate equations solved by Runge-Kutta algorithm. 
We used measured dispersion values for $\beta_{2}$, and calculated nonlinear coefficients $\gamma$=3.6$\times$ 10$^{-3}$ W$^{-1}$m$^{-1}$ and $1.3\times$ 10$^{-3}$ W$^{-1}$m$^{-1}$  for EDF and SMF, respectively. The saturable absorber (SA) was modelled by the following instantaneous transfer function:
 $P_o= T \,P_{i}$, where $T\equiv T_{0}+\frac{\Delta T P_{i}}{P_{i}+P_{sat}}$, 
describes the transmission of the SA, $P_i$  ($P_o$) being the instantaneous input (output) optical power, normalized as $P(z,t)\equiv |\psi (z,t)|^2$. 
As typical values, we took $T_{0}$=0.70 for the transmissivity at low signal and $\Delta T$=0.30 as the absorption contrast. We adjusted the mode-locking conditions by tuning the saturation power $P_{sat}$, and the pump power $P_{P,num}$. We obtained two 0.45 ps FWHM pulses separated by 1.2-2.0 ps, in good agreement with experiments. 

The vibrating soliton pair molecule is achieved for $P_{sat}$=9 W and $P_{P,num}$=42.5 mW. We also obtain the soliton pair molecules that mainly experience a sliding relative phase, in the form of oscillations ($P_{sat}$=10 W, $P_{P,num}$=47.5 mW) or infinite divergence ($P_{sat}$=16 W, $P_{P,num}$=47.5 mW). 
The numerical results are displayed on Fig.\ref{fig:6}. They reproduce qualitatively well our experimental observations. Panels (a), (b) and (d) of Fig.\ref{fig:6} show the spectral profiles as a function of cavity round trips for the three types of investigated soliton molecule dynamics. 
Panels (c) and (e) of Fig.\ref{fig:6} display the evolutions of $\tau$ and $\phi$, respectively, whereas Fig.\ref{fig:6}(f) plots the intensity of leading and trailing pulses versus cavity round trips. 

Numerical simulations present an additional feature, in the form of energy exchange among the two interacting solitons that becomes important in the phase-dominated dynamics of closely-spaced pulses. This feature was also identified numerically in Ref.\cite{Zav09}, whereas it may not occur for larger pulse spacing \cite{Gra06,Sot07}. 
When the leading pulse is larger than the trailing pulse, the difference of the phase velocities of leading and trailing pulses is negative, so that the phase difference decreases, and vice versa. When the intensities become equal, this produces the turning points of the oscillatory phase dynamics. 
In the case of a soliton molecule with diverging phase evolution, the intensity of one of the two pulses remains larger during propagation (Fig.\ref{fig:6}(f), black curve), so that the relative phase will keep on decreasing (increasing), as clearly shown in Fig.\ref{fig:6}(e) (black curve). 
This energy exchange feature illustrates the great interest of numerical simulation, which provide additional insight to the experimental characterization. We note that the retrieval of low-contrast pulse amplitude oscillation remains currently beyond the precision of our measurement technique, and is challenging future real-time measurements. 

To summarize, in this Letter we reported the first direct observation of dissipative optical soliton molecular motions, through real-time, single-shot spectral measurements allowed by the DFT method. We used a 1.55 $\mu$m fiber laser setup mode-locked through the virtual, ultrafast, NPE saturable absorber in the anomalous path-averaged dispersion regime. By tuning cavity parameters, the laser generated, in a reproducible way, soliton pair molecules of markedly different internal dynamics, from stationary to pulsating ones. 
Basing our characterization on the major degrees of freedom, namely the relative temporal separation and phase between the two bound solitons, we could evidence oscillating and vibrating soliton pairs, as well as a diverging phase dynamics, confirming the existence of internal dynamics previously anticipated. 
Our numerical simulations confirmed these typical signatures, while providing at the same time additional insight and incentive for future investigations, highlighting energy exchanges internal to the soliton molecule. 
We believe that our results have provided an additional building block for an appealing analogy between soliton molecules and molecules of matter, which should stimulate additional research of the self-organization in dissipative complex optical systems. 


\begin{acknowledgments}
We acknowledge support from CEFIPRA (project 5104$-$2), LABEX Action and R\'egion Bourgogne. K.N. acknowledges funding from CNRS. 
\end{acknowledgments}

 

\end{document}